\documentclass[conference,10pt]{IEEEtran}
\usepackage{cite}
\usepackage[cmex10]{amsmath}
\usepackage{array}
\usepackage{amsfonts}
\usepackage{mathrsfs}
\usepackage{arydshln}
\usepackage{slashbox}
\usepackage{graphicx}
\usepackage{float}
\usepackage{subfigure}
\usepackage{amssymb}
\usepackage{amsmath}
\usepackage[colorlinks,linkcolor=black,urlcolor=black,anchorcolor=black,citecolor=black,hyperfootnotes=true]{hyperref}
\usepackage{multirow}
\usepackage{multicol}
\usepackage{cite}
\usepackage[subfigure]{graphfig}
\usepackage{xcolor}
\usepackage{setspace}
\newcommand{\mv}[1]{\mbox{\boldmath{$ #1 $}}}
\usepackage[linesnumbered,ruled]{algorithm2e}
\newtheorem{theorem}{\underline{Theorem}}
\newtheorem{assumption}{\underline{Assumption}}
\newtheorem{lemma}{\underline{Lemma}}
\newtheorem{proposition}{\underline{Proposition}}

\newcommand{\qed}{\nobreak \ifvmode \relax \else
      \ifdim\lastskip<1.5em \hskip-\lastskip
      \hskip1.5em plus0em minus0.5em \fi \nobreak
      \vrule height0.75em width0.5em depth0.25em\fi}
\newcommand{\cc}{\mathrm{C}}
\newcommand{\dd}{\mathrm{D}}

\newcommand{\II}{\mathrm{I}}
\newcommand{\ub}{\mathrm{U}}
\begin{document}
\title{{Intelligent Reflecting Surface Aided Multiple Access: Capacity Region and Deployment Strategy}
\author{\IEEEauthorblockN{Shuowen~Zhang and Rui~Zhang}
	\IEEEauthorblockA{ECE Department, National University of Singapore. Email: \{elezhsh,elezhang\}@nus.edu.sg}\\[-4mm] \emph{(Invited Paper)}}\vspace{-3mm}}
\maketitle

\vspace*{-6mm}

\begin{abstract}
Intelligent reflecting surface (IRS) is a new promising technology that is able to manipulate the wireless propagation channel via smart and controllable signal reflection. In this paper, we investigate the \emph{capacity region} of a multiple access channel (MAC) with two users sending independent messages to an access point (AP), aided by $M$ IRS reflecting elements. We consider two practical IRS deployment strategies that lead to different user-AP effective channels, namely, the \emph{distributed deployment} where the $M$ reflecting elements form two IRSs, each deployed in the vicinity of one user, versus the \emph{centralized deployment} where all the $M$ reflecting elements are deployed in the vicinity of the AP. For the distributed deployment, we derive the capacity region in closed-form; while for the centralized deployment, we derive a capacity region outer bound and propose an efficient \emph{rate-profile} based method to characterize an achievable rate region (or capacity region inner bound). Furthermore, we compare the capacity regions of the two cases and draw useful insights into the optimal deployment of IRS in practical systems.
\end{abstract}
\vspace{-2mm}
\section{Introduction}
\vspace{-2mm}
Driven by the recent advancement in metamaterial technology, \emph{intelligent reflecting surface (IRS)} has become a cost-effective and energy-efficient solution to improve the wireless communication performance \cite{Towards,Survey_Basar}. Specifically, an IRS is a planar metasurface consisting of a large number of passive reflecting elements, each of which is able to introduce an independent phase shift to the impinging electromagnetic wave, thereby collaboratively altering the wireless channel. By properly designing the IRS reflection coefficients (i.e., phase shifts), IRS has been shown effective in improving the achievable rate of various wireless communication systems (see, e.g., \cite{Joint_Active,Protocol,ICASSP_CW,Emil}). Moreover, efficient channel estimation methods have been proposed to obtain the channel state information (CSI) required to practically realize the above rate gains \cite{CE_Johansson,Protocol,CE_B}. Nonetheless, from an information theoretical viewpoint, the fundamental \emph{capacity limit} of IRS-aided channels has only been characterized recently in \cite{MIMO} under the single-user setup. To the best of our knowledge, the \emph{capacity region} characterization for the more complex IRS-aided \emph{multi-user} channels still remains an open problem.

\begin{figure}[t]
	\centering
	\subfigure[Distributed deployment]{
		\includegraphics[width=4.1cm]{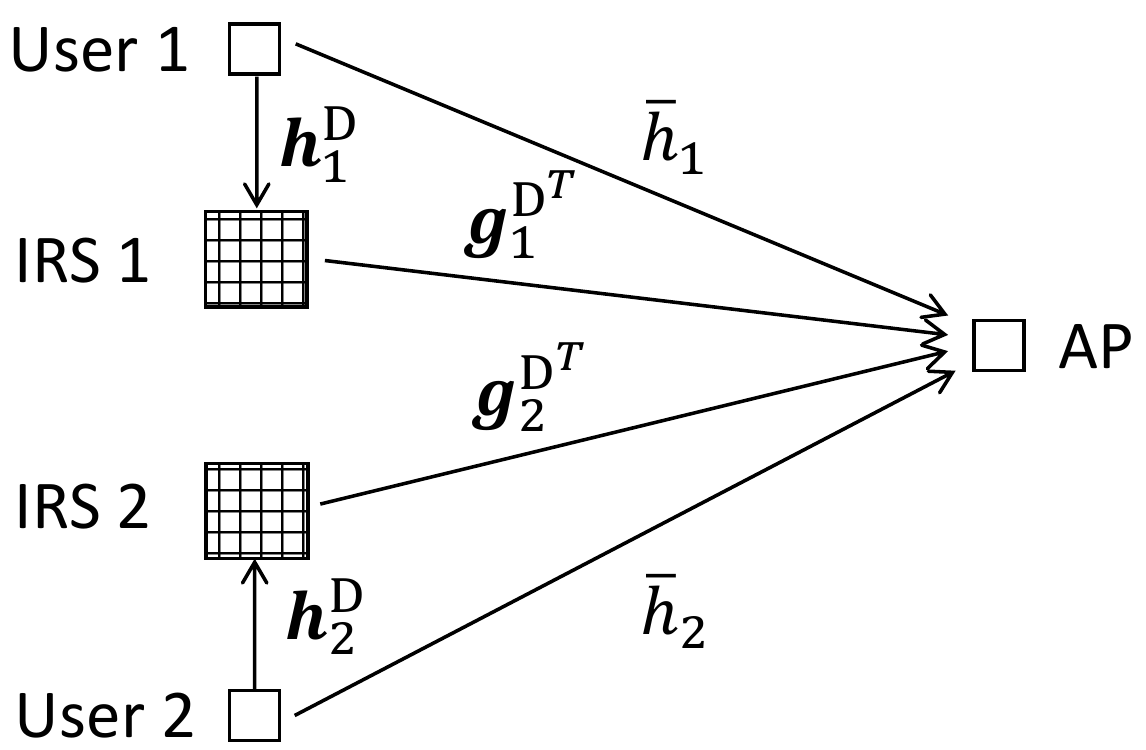}}
	\subfigure[Centralized deployment]{
		\includegraphics[width=4.1cm]{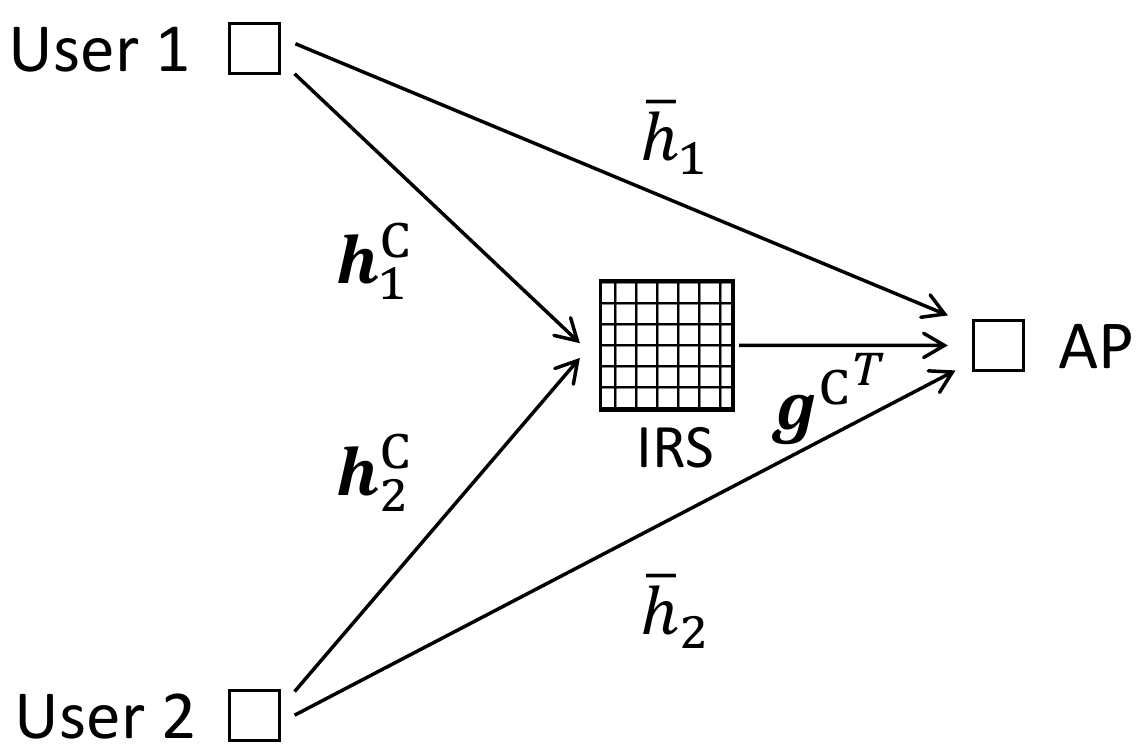}}
	
	\vspace{-3mm}
	
	\caption{A two-user MAC with different IRS deployment strategies.}\label{Fig_System}
	\vspace{-7.5mm}
\end{figure}

Besides capacity characterization, another key problem not fundamentally understood for IRS-aided multi-user systems is \emph{IRS deployment}. In the current literature, IRS is typically assumed to be deployed in the vicinity of the users to enhance the local signal coverage. Under this strategy, multiple distributed IRSs need to be deployed each near one cluster of users if the users in different clusters are located far apart, which is referred to as the \emph{distributed deployment} and illustrated in Fig. \ref{Fig_System} (a). In contrast, given a total number of available IRS reflecting elements, another strategy is the \emph{centralized deployment} where all reflecting elements are deployed near the access point (AP), as illustrated in Fig. \ref{Fig_System} (b). Note that these two strategies lead to different user-AP effective channels in general and hence different user achievable rates. Specifically, with distributed deployment, each user can only enjoy the passive beamforming gain brought by its nearby IRS (since its signals reflected by other far-apart IRSs are too weak due to much higher path loss), which is thus smaller than the passive beamforming gain under the centralized deployment with a larger-size IRS where all the reflecting elements can be used for enhancing the channels for all users. However, the IRS passive beamforming gain under the centralized deployment needs to be shared by all users, thus resulting in a reduced gain for each user. To our best knowledge, it is yet unclear which IRS deployment strategy achieves larger capacity region in multi-user systems.

To address the above issue, we study in this paper a two-user multiple access channel (MAC) aided by $M$ IRS reflecting elements, as shown in Fig. \ref{Fig_System}. For the distributed IRS deployment, we provide a \emph{closed-form} characterization of its capacity region. While for the centralized IRS deployment, we propose a capacity region outer bound and develop a computationally efficient \emph{rate-profile} based method to characterize an achievable rate region (or capacity region inner bound). Moreover, we analytically prove that the capacity region with centralized deployment contains that with distributed deployment under a simplified but practical setup. Numerical results validate our analysis and tightness of the proposed bounds. Furthermore, it is shown that the capacity gain of centralized over distributed deployment is most prominent when the rates of the two users are \emph{asymmetric}.
\vspace{-2mm}
\section{System Model}
\vspace{-1mm}
We consider a two-user MAC in Fig. \ref{Fig_System}, where each single-antenna user aims to send an independent message to a single-antenna AP. The baseband equivalent direct channel from the $k$th user to the AP is denoted as $\bar{h}_k\in \mathbb{C}$, $k=1,2$. To improve the user communication rates, we consider the deployment of $M\geq 1$ passive reflecting elements, each element being able to induce an independent phase shift to the incident signal, thus collaboratively altering the effective channels from the users to the AP. We propose two different deployment strategies for the $M$ reflecting elements. Specifically, for the \emph{distributed deployment}, the $M$ elements form two IRSs (see Fig. \ref{Fig_System} (a)), where each IRS $k$ consists of $M_k$ elements and is placed in the vicinity of user $k$, with $\sum_{k=1}^2 M_k=M$. In contrast, for the \emph{centralized deployment}, all the $M$ elements form one single IRS located in the vicinity of the AP (see Fig. \ref{Fig_System} (b)). In the following, we describe the system models for the two deployment cases, respectively.
\vspace{-3mm}
\subsection{Distributed IRS Deployment}
\vspace{-1mm}
For distributed IRS deployment, we denote ${\mv{h}}_k^\dd\in \mathbb{C}^{M_k\times 1}$ as the channel vector from user $k$ to its serving IRS, and ${\mv{g}}_k^{\dd^T}\in \mathbb{C}^{1\times M_k}$ as the channel vector from its serving IRS to the AP. Denote ${\mv{\Phi}}^\dd_k=\mathrm{diag}\{\phi^\dd_{k1},...,\phi^\dd_{k M_k}\}\in \mathbb{C}^{M_k\times M_k}$ as the IRS reflection matrix for the $k$th IRS, with $|\phi^\dd_{km}|=1,\ \forall m\in {\mathcal{M}}_k$, where ${\mathcal{M}}_k=\left\{1,...,M_k\right\}$. We assume that the locations of the two users are sufficiently far apart such that the signal transmitted by one user and reflected by the other user's serving IRS is negligible at the AP due to the high path loss. Hence, the effective channel from user $k$ to the AP by combining both the direct and reflected links is given by
\vspace{-2mm}
\begin{equation}\label{channel_distributed}
\tilde{h}^\dd_k(\mv{\Phi}_k^\dd)=\bar{h}_k+{\mv{g}}_k^{\dd^T}{\mv{\Phi}}^\dd_k{\mv{h}}_k^\dd,\quad k=1,2.
\vspace{-2mm}\end{equation}
Let $s_k$ denote the desired information symbol for user $k$, which is assumed to be a circularly symmetric complex Gaussian (CSCG) random variable with zero mean and unit variance, i.e., $s_k\sim \mathcal{CN}(0,1)$. Note that $s_k$'s are independent over $k$. The transmitted signal by user $k$ is modeled as $x_k=\sqrt{p_k}s_k$, which satisfies $\mathbb{E}[|x_k|^2]=p_k\leq P_k$, with $p_k$ denoting the transmit power of user $k$ and $P_k$ denoting its maximum value. The received signal at the AP is thus modeled as
\vspace{-2mm}\begin{equation}\label{received_distributed}
y=\tilde{h}^\dd_1(\mv{\Phi}_1^\dd)x_1+\tilde{h}^\dd_2(\mv{\Phi}_2^\dd)x_2+z,
\vspace{-2mm}\end{equation}
where $z\!\sim\! \mathcal{CN}(0,\sigma^2)$ denotes the CSCG noise at the AP receiver with average power $\sigma^2$. For each user $k$, we let $R^\dd_k$ denote its achievable rate in bits per second per Hertz (bps/Hz) under the distributed IRS deployment.
\vspace{-3mm}
\subsection{Centralized IRS Deployment}
\vspace{-1mm}
For centralized IRS deployment, we denote $\mv{h}_k^\cc\in \mathbb{C}^{M\times 1}$ as the channel vector from user $k$ to the IRS, and $\mv{g}^{\cc^T}\in \mathbb{C}^{1\times M}$ as the channel vector from the IRS to the AP. Denote $\mv{\Phi}^\cc=\mathrm{diag}\{\phi_1^\cc,...,\phi_M^\cc\}\in \mathbb{C}^{M\times M}$ as the IRS reflection matrix, with $|\phi^\cc_m|=1,\ \forall m\in \mathcal{M}$, where $\mathcal{M}=\{1,...,M\}$. Thus, the effective channel from user $k$ to the AP is given by
\vspace{-2mm}
\begin{equation}
\tilde{h}^\cc_k(\mv{\Phi}^\cc)=\bar{h}_k+\mv{g}^{\cc^T}\mv{\Phi}^\cc\mv{h}_k^\cc,\quad k=1,2.
\vspace{-2mm}
\end{equation}
Note that different from the distributed deployment where the effective channel between each user $k$ and the AP is only dependent on the $M_k$ reflection coefficients of its own serving IRS in $\mv{\Phi}_k^\dd$, the effective channels for both users under the centralized deployment depend on all the $M$ reflection coefficients in $\mv{\Phi}^\cc$. Under the same transmitted signal and receiver noise model as in the distributed deployment case, the received signal at the AP is modeled similarly as (\ref{received_distributed}) by replacing each $\tilde{h}^\dd_k(\mv{\Phi}_k^\dd)$ with $\tilde{h}^\cc_k(\mv{\Phi}^\cc)$. For each user $k$, let $R^\cc_k$ denote the achievable rate in bps/Hz under the centralized IRS deployment.

In this paper, we aim to characterize the \emph{capacity region} of the IRS-aided two-user MAC under the two deployment strategies, namely, all the achievable rate-pairs $(R^\dd_1,R^\dd_2)$'s and $(R^\cc_1,R^\cc_2)$'s. We then compare these two capacity regions and draw useful insights on the optimal IRS deployment strategy.
\vspace{-1mm}
\section{Capacity Region of Distributed Deployment}
\vspace{-1mm}
First, we characterize the capacity region under the distributed IRS deployment. Note that with given IRS reflection coefficients $\{{\mv{\Phi}}^\dd_k\}$, the channels from the two users to the AP are determined as $\{\tilde{h}_k^\dd(\mv{\Phi}_k^\dd)\}$ given in (\ref{channel_distributed}), and the capacity region of the two-user MAC is well-known as the convex hull of all rate-pairs that satisfy the following constraints \cite{Elements}:
\vspace{-2mm}\begin{align}
{R}_1^\dd\leq &\log_2(1+{P_1|{\tilde{h}}^\dd_1(\mv{\Phi}_1^\dd)|^2}/{\sigma^2})\overset{\Delta}{=}r^\dd_1({\mv{\Phi}}^\dd_1),\label{D1}\\[-1mm]
{R}_2^\dd\leq &\log_2(1+{P_2|{\tilde{h}}^\dd_2(\mv{\Phi}_2^\dd)|^2}/{\sigma^2})\overset{\Delta}{=}r^\dd_2({\mv{\Phi}}^\dd_2),\label{D2}
\end{align}
\vspace{-6mm}\begin{equation}
\!\!\!\!{R}_1^\dd\!+\!{R}_2^\dd\!\leq\! \log_2(1\!+\!{\textstyle\sum}_{k=1}^2P_k|{\tilde{h}}^\dd_k(\mv{\Phi}_k^\dd)|^2/\sigma^2)\!\overset{\Delta}{=}\!r^\dd_{12}(\{{\mv{\Phi}}^\dd_k\}),\label{D3}
\vspace{-1mm}\end{equation}
which is denoted as ${\mathcal{C}}^\dd(\{{\mv{\Phi}}^\dd_k\})$. Note that by flexibly designing the IRS reflection coefficients $\{\mv{\Phi}^\dd_k\}$, any rate-pair within the union set of ${\mathcal{C}}^\dd(\{{\mv{\Phi}}^\dd_k\})$'s over all feasible $\{\mv{\Phi}^\dd_k\}$'s can be achieved. By further considering \emph{time sharing} among different $\{\mv{\Phi}^\dd_k\}$'s, the capacity region under distributed IRS deployment is defined as the convex hull of such a union set \cite{Elements}:
\vspace{-2mm}\begin{equation}\label{Cdis}
\mathcal{C}^\dd\overset{\Delta}{=}\mathrm{Conv}\Big({\bigcup}_{\{{\mv{\Phi}}^\dd_k\}\in \mathcal{R}^\dd} {\mathcal{C}}^\dd(\{{\mv{\Phi}}^\dd_k\})\Big),
\vspace{-3mm}\end{equation}
where $\mathrm{Conv}(\cdot)$ denotes the convex hull operation, and $\mathcal{R}^\dd\!\overset{\Delta}{=}\!\{{\{{\mv{\Phi}}^\dd_k\}\!:\!|\phi^\dd_{km}|\!=\!1,\forall k, m}\}$ denotes the feasible set of ${\{{\mv{\Phi}}^\dd_k\}}$. 

In the following, we characterize $\mathcal{C}^\dd$ in closed-form by exploiting the peculiar effective channel structure under the distributed deployment. Specifically, note that for any $\{\mv{\Phi}_k^\dd\}\!\in\! \mathcal{R}^\dd$, the effective channel gain for each user $k$ is upper-bounded by
\vspace{-0mm}\begin{align}\label{eq1}
|\tilde{h}_k^\dd(\mv{\Phi}_k^\dd)|\!=&|\bar{h}_k\!\!+\!\!{\textstyle\sum}_{m=1}^{M_k}{g}^\dd_{km}{\phi}^\dd_{km}{h}_{km}^\dd|\!\leq\! |\bar{h}_k|\!\!+\!\!{\textstyle\sum}_{m=1}^{M_k}|g^\dd_{km}||h_{km}^\dd|\nonumber\\[-1mm]
=&
|\bar{h}_k|+\|\mathrm{diag}\{{\mv{g}}^\dd_k\}{\mv{h}}_k^\dd\|_1\overset{\Delta}{=}\tilde{h}_{k,\ub}^\dd,\ k=1,2,
\end{align}
where $\|\cdot\|_1$ denotes the $l_1$-norm, and the inequality holds with equality if and only if $\{\mv{\Phi}_k^\dd\}$ is designed as follows:
\vspace{-2mm}\begin{equation}\label{eq2}
\phi_{km}^\dd=e^{j(\arg\{\bar{h}_k\}-\arg\{g^\dd_{km}h_{km}^\dd\})},\quad k=1,2,\ m\in \mathcal{M}_k.
\vspace{-2mm}\end{equation}
Based on this result, we obtain the following theorem.
\begin{theorem}\label{theorem_dis}
The capacity region of the IRS-aided two-user MAC under the distributed deployment is given by
\vspace{-2mm}\begin{equation}\label{C_dis}
\!\!\mathcal{C}^\dd\!\!=\!\!\{(R^\dd_1,R^\dd_2)\!:\!R^\dd_1\!\leq\! r_1^{\dd^\star}\!,R^\dd_2\!\leq\! r_2^{\dd^\star}\!,R^\dd_1\!+\!R^\dd_2\!\leq \!r_{12}^{\dd^\star}\!\},
\vspace{-2mm}\end{equation}
where $r_1^{\dd^\star}\!\overset{\Delta}{=}\!\log_2(1\!+\!{P_1\tilde{h}_{1,\ub}^{\dd^2}}/{\sigma^2})$, $r_2^{\dd^\star}\!\overset{\Delta}{=}\!\log_2(1\!+\!{P_2\tilde{h}_{2,\ub}^{\dd^2}}/{\sigma^2})$, and $r_{12}^{\dd^\star}\overset{\Delta}{=}\log_2\big(1+(P_1\tilde{h}_{1,\ub}^{\dd^2}+P_2\tilde{h}_{2,\ub}^{\dd^2})/\sigma^2\big)$.
\end{theorem}
\begin{IEEEproof}[Proof (Sketch)]
Theorem \ref{theorem_dis} can be proved by noting that (\ref{C_dis}) is an achievable rate region with $\{\mv{\Phi}_k^\dd\}$ given in (\ref{eq2}), and also a convex-shaped outer bound for all achievable ${\mathcal{C}}^\dd(\{{\mv{\Phi}}^\dd_k\})$'s (thus, the convex-hull operation in (\ref{Cdis}) is not needed with $\{{\mv{\Phi}}^\dd_k\}$ given in (\ref{eq2})).
\end{IEEEproof}
\vspace{-1.5mm}
\section{Capacity Region of Centralized Deployment}
\vspace{-1.5mm}
Next, we characterize the capacity region under the centralized IRS deployment. Similar to the distributed case, the capacity region with given IRS reflection coefficients ${\mv{\Phi}}^\cc$ is the convex hull of the rate-pairs under the following constraints:
\vspace{-6mm}\begin{align}
{R}_1^\cc\leq &\log_2(1+P_1|{\tilde{h}}^\cc_1(\mv{\Phi}^\cc)|^2/\sigma^2)\overset{\Delta}{=}r^\cc_1(\mv{\Phi}^\cc),\label{C1}\\[-1mm]
{R}_2^\cc\leq &\log_2(1+P_2|{\tilde{h}}^\cc_2(\mv{\Phi}^\cc)|^2/\sigma^2)\overset{\Delta}{=}r^\cc_2(\mv{\Phi}^\cc),\label{C2}
\end{align}
\vspace{-6mm}
\begin{equation}
{R}_1^\cc\!+\!{R}_2^\cc\!\leq\! \log_2(1\!+\!{\textstyle\sum}_{k=1}^2P_k|{\tilde{h}}^\cc_k({\mv{\Phi}}^\cc)|^2/\sigma^2)\!\overset{\Delta}{=}\!r^\cc_{12}(\mv{\Phi}^\cc),\label{C3}
\vspace{-2mm}\end{equation}
which is denoted as ${\mathcal{C}}^\cc({\mv{\Phi}}^\cc)$. By tuning the IRS reflection coefficients $\mv{\Phi}^\cc$ and performing time sharing among different $\mv{\Phi}^\cc$'s, the capacity region is defined as
\vspace{-2mm}\begin{equation}\label{C_cen}
\mathcal{C}^\cc\overset{\Delta}{=}\mathrm{Conv}\Big({\bigcup}_{{\mv{\Phi}}^\cc\in \mathcal{R}^\cc} {\mathcal{C}}^\cc({\mv{\Phi}}^\cc)\Big),
\vspace{-3mm}\end{equation}
where $\mathcal{R}^\cc\overset{\Delta}{=}\{{\mv{\Phi}}^\cc:|\phi^\cc_{m}|=1,\forall m\}$ denotes all feasible ${\mv{\Phi}}^\cc$'s.

Compared to the distributed deployment case, the capacity region in (\ref{C_cen}) is more challenging to characterize. This is because the effective channels of the two users, $\tilde{h}_1^\cc(\mv{\Phi}^\cc)$ and $\tilde{h}_2^\cc(\mv{\Phi}^\cc)$, are \emph{coupled} through all the $M$ reflection coefficients in $\mv{\Phi}^\cc$. Thus, different parts of the Pareto boundary of the capacity region $\mathcal{C}^\cc$ are generally achieved by different $\mv{\Phi}^\cc$ to strike a balance between $\tilde{h}_1^\cc(\mv{\Phi}^\cc)$ and $\tilde{h}_2^\cc(\mv{\Phi}^\cc)$. Finding such capacity-achieving sets of reflection coefficients is more challenging as compared to the distributed case where the entire Pareto boundary of the capacity region is achieved by a single set of $\{\mv{\Phi}_k^\dd\}$ given in (\ref{eq2}), since the effective channel of each user is maximized by the reflection coefficients of its own serving IRS. Although $\mathcal{C}^\cc$ can be characterized via the \emph{exhaustive search} by first obtaining $\mathcal{C}^\cc(\mv{\Phi}^\cc)$'s for all feasible ${\mv{\Phi}}^\cc\!\!\in\!\! \mathcal{R}^\cc$ and then taking the convex hull of their union set, the required complexity is at least $\mathcal{O}(L_0^M)$ if the $[0,2\pi)$ phase range for each $\phi_m^\cc$ in ${\mv{\Phi}}^\cc$ is approximated by $L_0$ uniformly sampled points, which is \emph{exponential} over $M$ and thus prohibitive for practically large $M$. To avoid such high complexity for characterizing $\mathcal{C}^\cc$, in the following, we provide efficient methods to find both the outer and inner bounds of $\mathcal{C}^\cc$, whose tightness will be evaluated via numerical results in Section \ref{sec_num}.
\vspace{-3mm}
\subsection{Capacity Region Outer Bound}
\vspace{-1mm}
To start with, we provide an outer bound of the capacity region $\mathcal{C}^\cc$. Specifically, it follows from (\ref{C1})--(\ref{C3}) that an outer bound of $\mathcal{C}^\cc$ can be constructed by finding an upper bound for each of $r_1^\cc(\mv{\Phi}^\cc)$, $r_2^\cc(\mv{\Phi}^\cc)$, and $r_{12}^\cc(\mv{\Phi}^\cc)$ separately, for which the details are given as follows.

First, similar to (\ref{eq1}), it can be shown that for each user $k$, the effective channel gain $|\tilde{h}^{\cc}_k(\mv{\Phi}^\cc)|$ is upper-bounded by
\vspace{-2mm}\begin{equation}
|\tilde{h}^{\cc}_k(\mv{\Phi}^\cc)|\leq |\bar{h}_k|+\|\mathrm{diag}\{\mv{g}^\cc\}\mv{h}_k^\cc\|_1\overset{\Delta}{=}\tilde{h}_{k,\ub}^{\cc},
\vspace{-2mm}\end{equation}
where the inequality holds with equality if and only if all the IRS reflection coefficients are designed to maximize user $k$'s effective channel gain, i.e.,
\vspace{-2mm}\begin{equation}\label{phim}
\phi_{m}^\cc=e^{j(\arg\{\bar{h}_k\}-\arg\{g_m^\cc h_{km}^\cc\})},\quad 
m\in \mathcal{M}.
\vspace{-2mm}\end{equation}
Thus, based on (\ref{C1})--(\ref{C2}), each $r_k^\cc(\mv{\Phi}^\cc)$ is upper-bounded as
\vspace{-3mm}
\begin{equation}
r_k^\cc(\mv{\Phi}^\cc)\leq \log_2(1+P_k\tilde{h}_{k,\ub}^{{\cc}^2}/\sigma^2)\overset{\Delta}{=}r_{k,\ub}^{\cc},\quad k=1,2.
\vspace{-2mm}\end{equation}

Next, we derive an upper bound for $r_{12}^\cc(\mv{\Phi}^\cc)$, which is a challenging task since $\mv{\Phi}^\cc$ can change both $\tilde{h}^{\cc}_1({\mv{\Phi}}^\cc)$ and $\tilde{h}^{\cc}_2(\mv{\Phi}^\cc)$ in $r_{12}^\cc(\mv{\Phi}^\cc)$. To achieve this goal, we formulate the following optimization problem:
\vspace{-3mm}\begin{equation}
\mbox{(P0)} \underset{\mv{\Phi}^\cc:|\phi_m^\cc|=1,\forall m\in \mathcal{M}}{\mathtt{max}} P_1|\tilde{h}_1^\cc({\mv{\Phi}}^\cc)|^2+P_2|\tilde{h}_2^\cc({\mv{\Phi}}^\cc)|^2.
\vspace{-2mm}\end{equation} 
Let $s_0^\star$ denote the optimal value of (P0). Note that for any $s_0\!\geq\! s_0^\star$, $\log_2(1\!+\!s_0/\sigma^2)$ is an upper bound for $r_{12}^\cc(\mv{\Phi}^\cc)$. However, (P0) is a non-convex optimization problem due to the uni-modular constraints on $\phi_m^\cc$'s, thus $s_0^\star$ is generally difficult to obtain. In the following, we find an upper bound for $s_0^\star$ instead.

First, we transform (P0) into a more tractable form. Define ${\mv{q}}_k^H\overset{\Delta}{=}{\mv{g}}^{\cc^T}\mathrm{diag}\{\mv{h}_k^\cc\}$, $\mv{v}\overset{\Delta}{=}P_1\bar{h}_1\mv{q}_1+P_2\bar{h}_2\mv{q}_2$, and $\mv{\phi}^\cc\overset{\Delta}{=}[\phi_1^\cc,...,\phi_M^\cc]^T$. Consequently, the objective function of (P0) can be rewritten as $P_1|\tilde{h}_1^\cc({\mv{\Phi}}^\cc)|^2+P_2|\tilde{h}_2^\cc({\mv{\Phi}}^\cc)|^2=P_1|\bar{h}_1|^2+P_2|\bar{h}_2|^2+{\mv{v}}^H\mv{\phi}^\cc
+{\mv{\phi}}^{\cc^H}\mv{v}
+\mv{\phi}^{\cc^H}(P_1\mv{q}_1\mv{q}_1^H+P_2\mv{q}_2\mv{q}_2^H)\mv{\phi}^\cc$, 
which is a quadratic function of $\mv{\phi}^\cc$. Thus, we can apply the semidefinite relaxation (SDR) technique for finding an upper bound for the optimal value of (P0). By introducing auxiliary variables ${\mv{w}}=[\mv{\phi}^{\cc^T},t]^T$ and $\mv{W}=\mv{ww}^H$, (P0) can be shown to be equivalent to the following problem with an additional constraint of $\mathrm{rank}(\mv{W})=1$:
\vspace{-2.5mm}\begin{equation}
\!\!\mbox{(P0-SDR)} \underset{\scriptstyle \small{\mv{W}}\succeq\mv{0}:\small{\mv{W}}_{m,m}=1,\atop\scriptstyle m=1,...,M+1}{\mathtt{max}} P_1|\bar{h}_1|^2+P_2|\bar{h}_2|^2+\mathrm{tr}\{\mv{WQ}\},\!\!
\vspace{-3mm}\end{equation} 
where $\mv{Q}\overset{\Delta}{=}[P_1\mv{q}_1\mv{q}_1^H+P_2\mv{q}_2\mv{q}_2^H,\mv{v};\mv{v}^H,0]$.
(P0-SDR) is a semidefinite program (SDP) which can be efficiently solved via the interior-point method with complexity $\mathcal{O}(M^{4.5})$ \cite{SDR}. Denote $s^\star$ as the optimal value of (P0-SDR). Note that $s^\star\geq s_0^\star$ holds due to the relaxation of the rank-one constraint. Therefore, we have $r_{12}^\cc(\mv{\Phi}^\cc)\leq \log_2(1+s^\star/\sigma^2)\overset{\Delta}{=}r_{12,\ub}^{\cc}$, which yields an outer bound of ${\mathcal{C}}^\cc$ given by
\vspace{-2mm}\begin{equation}\label{OB2}
\!\!{\mathcal{C}}^{\cc}_{\mathrm{O}}\!\!=\!\!\{(R_1^\cc,R_2^\cc)\!:\!R_1^\cc\!\leq\! r_{1,\ub}^{\cc},R_2^\cc\!\leq\! r_{2,\ub}^{\cc},R_1^\cc\!+\!R_2^\cc\!\leq\! r_{12,\ub}^{\cc}\}\!\supseteq\!{\mathcal{C}}^\cc.
\end{equation}
\vspace{-8mm}
\subsection{Capacity Region Inner Bound: A Rate-Profile Method}
\vspace{-1mm}
Next, we derive an inner bound of the capacity region $\mathcal{C}^\cc$ (or an achievable rate region). We first present a \emph{rate-profile} based method to achieve this goal by solving a series of sum-rate maximization problems. Then, we propose an \emph{alternating optimization} algorithm to find high-quality solutions to these problems efficiently.
\subsubsection{Rate-Profile based Problem Formulation}
To start with, note that for each $\mv{\Phi}^\cc$, all the achievable rate-pairs on the Pareto boundary of its corresponding $\mathcal{C}^\cc(\mv{\Phi}^\cc)$ except those requiring time sharing/rate splitting of the two users can be attained via \emph{successive interference cancellation (SIC)} at the AP, i.e., first decoding the message of one user by treating the signal of the other user as noise, then canceling the decoded signal and decoding the other user's message \cite{Elements}. Motivated by this result, we propose to first characterize the Pareto boundary of the union set of the above SIC-achievable rate-pairs for all feasible $\mv{\Phi}^\cc\!\in\! \mathcal{R}^\cc$, and then perform time sharing among the obtained rate-pairs on the Pareto boundary to further enlarge the achievable rate region. For the first task, we propose to adopt the \emph{rate-profile} approach in \cite{Cooperative}. Specifically, let $\mv{\pi}$ denote the decoding order indicator, with $\mv{\pi}\!=\![1,2]^T\!\overset{\Delta}{=}\!\mv{\pi}^{\II}$ representing that user $1$ is decoded before user $2$, and $\mv{\pi}\!=\![2,1]^T\!\overset{\Delta}{=}\!\mv{\pi}^{\II\II}$ otherwise; let $\alpha\!\in\! [0,1]$ denote the rate ratio between the firstly decoded user and the users' sum-rate, and $\mv{\alpha}\!=\![\alpha,1-\alpha]^T$ denote the rate-profile vector. Based on the above, we formulate the following problem to maximize the sum-rate of the two users with given $\mv{\alpha}$ and $\mv{\pi}$ by jointly optimizing \hbox{the IRS reflection coefficients and user transmit powers:}
\vspace{-2mm}
\begin{align}
\!\!\!\mbox{(P1)} \underset{r,p_1,p_2,\mv{\Phi}^\cc}{\mathtt{max}} &r\\[-4mm]
\mathtt{s.t.}\quad &\log_2\bigg(1\!+\!\frac{p_{\pi_1}|\tilde{h}_{\pi_1}^\cc(\mv{\Phi}^\cc)|^2}{p_{\pi_2}|\tilde{h}_{\pi_2}^\cc(\mv{\Phi}^\cc)|^2\!+\!\sigma^2}\bigg)\!\geq\! \alpha r\label{P1c1}\\[-1mm]
&\log_2(1+p_{\pi_2}|\tilde{h}_{\pi_2}^\cc(\mv{\Phi}^\cc)|^2/\sigma^2)\geq (1-\alpha) r\label{P1c2}\\[-1mm]
& p_k\leq P_k,\quad  \forall k\in\{1,2\}\label{P1c3}\\[-1mm]
& \mv{\Phi}^\cc=\mathrm{diag}\{\phi_1^\cc,...,\phi_M^\cc\}\label{P1c4}\\[-1mm]
& |\phi_m^\cc|=1,\quad \forall m\in \mathcal{M}.\label{P1c5}
\end{align}

\vspace{-2mm}
For each rate-profile vector $\mv{\alpha}$, let $r_\II^\star(\mv{\alpha})$ and $r_{\II\II}^\star(\mv{\alpha})$ denote the optimal value to (P1) with $\mv{\pi}\!=\!\mv{\pi}^{\II}$ and $\mv{\pi}\!=\!\mv{\pi}^{\II\II}$, respectively. Note that $r_\II^\star(\mv{\alpha})\!\geq\! r_{\II\II}^\star(\mv{\alpha})$ represents that decoding order $\mv{\pi}^{\II}$ is optimal for the given $\mv{\alpha}$, and $r_\II^\star(\mv{\alpha})\!<\! r_{\II\II}^\star(\mv{\alpha})$ otherwise. Therefore, the Pareto-optimal rate-pair $(R_1^\cc,R_2^\cc)$ along the rate-profile vector $\mv{\alpha}$ is $(\alpha r_\II^\star(\mv{\alpha}),(1\!-\!\alpha)r_\II^\star(\mv{\alpha}))$ if $r_\II^\star(\mv{\alpha})\!\geq\! r_{\II\II}^\star(\mv{\alpha})$, and $((1-\alpha) r_{\II\II}^\star(\mv{\alpha}),\alpha r_{\II\II}^\star(\mv{\alpha}))$ otherwise.\footnote{It is worth noting that another approach to characterize the aforementioned Pareto boundary is by solving a series of \emph{weighted sum-rate maximization (WSRmax)} problems \cite{Cooperative}, which is also challenging since the rates of the two users are coupled in the objective function in a complicated manner. Thus, we leave the WSRmax-based approach to our future work.} In the following, we address the remaining \hbox{problem of solving (P1).} 
\subsubsection{Proposed Solution to (P1)}
\hspace{-2mm} Note that (P1) is a non-convex optimization problem due to the uni-modular constraints on $\phi_m^\cc$'s in (\ref{P1c5}), and the complicated coupling among $p_1$, $p_2$, and $\mv{\Phi}^\cc$ in (\ref{P1c1})--(\ref{P1c2}). To tackle such difficulty, we exploit the structure of (P1) to transform it into a more tractable form.
\vspace{-1mm}
\begin{proposition}\label{prop_P1}
(P1) is equivalent to the following problem:
\vspace{-6mm}
\begin{align}
\!\!\!	\mbox{(P2)}\ \underset{r,\mv{\Phi}^\cc}{\mathtt{max}}\ &r\\[-2mm]
	\mathtt{s.t.}\ &\log_2(1+{P_{\pi_1}|\tilde{h}_{\pi_1}(\mv{\Phi}^\cc)|^2}/({2^{(1-\alpha)r}\sigma^2}))\geq \alpha r\label{P2c1}\\[-1.5mm]
	&\log_2(1+{P_{\pi_2}|\tilde{h}_{\pi_2}(\mv{\Phi}^\cc)|^2}/{\sigma^2})\geq (1-\alpha) r\label{P2c2}\\[-1.5mm]
	& (\ref{P1c4}),(\ref{P1c5}).
	\end{align}
\end{proposition}
\vspace{-2mm}
\begin{IEEEproof}
Proposition \ref{prop_P1} can be proved by noting that the inequality in (\ref{P1c2}) can be replaced with equality without loss of optimality. We omit the details here for brevity.
\end{IEEEproof}

Note that for the case of $\alpha=1$, the optimal $\mv{\Phi}^\cc$ to (P2) can be readily derived as ${\phi}_{m}^\cc=e^{j(\arg\{\bar{h}_{\pi_1}\}-\arg\{{g}_{m}^\cc{h}_{{\pi_1}m}^\cc\})},\forall m$. Thus, we focus on (P2) with $\alpha\in [0,1)$ in the following. To further simplify (P2), we define an auxiliary variable $\beta\overset{\Delta}{=}2^{(1-\alpha)r}$, which is an increasing function of $r$ for any $\alpha\in [0,1)$. (P2) is then equivalently rewritten as
\vspace{-3mm}
\begin{align}
\mbox{(P3)}\quad \underset{\beta,\mv{\Phi}^\cc}{\mathtt{max}}\quad &\beta\\[-3mm]
\mathtt{s.t.}\quad &|\tilde{h}_{\pi_1}^\cc(\mv{\Phi}^\cc)|^2\geq (\beta^{\frac{1}{1-\alpha}}-\beta){\sigma^2}/P_{\pi_1}\label{P3c1}\\[-1mm]
&|\tilde{h}_{\pi_2}^\cc(\mv{\Phi}^\cc)|^2\geq (\beta-1){\sigma^2}/P_{\pi_2}\label{P3c2}\\[-1mm]
& (\ref{P1c4}),(\ref{P1c5}).
\end{align}

\vspace{-3mm}
(P3) is still non-convex due to the uni-modular constraints on $\phi_m^\cc$'s as well as the quadratic terms at the left-hand sides of (\ref{P3c1}) and (\ref{P3c2}), for which the optimal solution is thus difficult to obtain. In the following, we adopt an \emph{alternating optimization} approach for finding a high-quality suboptimal solution to (P3). Specifically, note that each quadratic term $|\tilde{h}_k^\cc(\mv{\Phi}^\cc)|^2$ can be expressed as the following \emph{affine} form over each $\phi_m^\cc$ with $\{\phi_i^\cc,i\neq m\}_{i=1}^M$ being fixed:
\vspace{-2mm}\begin{equation}\label{affine}
|\tilde{h}_k^\cc(\mv{\Phi}^\cc)|^2
=2\mathfrak{Re}\{f_{2,km}\phi_m^\cc\}\! +\!f_{1,km},\quad k=1,2,
\vspace{-2mm}\end{equation}
where $f_{1,km}\!\overset{\Delta}{=}\!|\bar{h}_k\!+\!\sum_{i\neq m} g_i^\cc\phi_i^\cc h_{ki}^\cc|^2\!+\!|g_m^\cc h_{km}^\cc|^2$ and $f_{2,km}\!\overset{\Delta}{=}\!g_m^\cc h_{km}^\cc({\bar{h}_k}^*\!+\!\sum_{i\neq m} g_i^{\cc^*}\phi_i^{\cc^*}h_{ki}^{\cc^*})$, and the equality in (\ref{affine}) holds due to $|\phi_m^\cc|\!=\!1$. Hence, with given $\{\phi_i^\cc,i\neq m\}_{i=1}^M$, (P3) is reduced to the following problem:
\vspace{-3mm}\begin{align}
\!\!\!\!\!\!\mbox{(P3-m)} \underset{\beta,\phi_m^\cc}{\mathtt{max}}\ &\beta\\[-2mm]
\mathtt{s.t.}\ &2\mathfrak{Re}\{f_{2,\pi_1m}\phi_m\}\!\geq\! (\beta^{\frac{1}{1-\alpha}}\!-\!\beta)\sigma^2\!/\!P_{\pi_1}\!\!-\!\!f_{1,\pi_1m}\label{P3mc1}\\[-1mm]
&2\mathfrak{Re}\{f_{2,\pi_2m}\phi_m\}\!\geq\! (\beta-1)\sigma^2\!/\!P_{\pi_2}\!\!-\!\!f_{1,\pi_2m}\label{P3mc2}\\[-1mm]
& |\phi_m^\cc|=1.\label{P3mc3}
\end{align}

\vspace{-2.5mm}
Note that the only non-convexity in (P3-m) lies in the uni-modular constraint on $\phi_m^\cc$, thus motivating us to apply the \emph{convex relaxation} technique on this constraint. Specifically, we relax (P3-m) by replacing the constraint in (\ref{P3mc3}) with a new convex constraint $|\phi_m^\cc|\leq 1$, and denote the relaxed problem as (P3-m-R). We then have the following lemma.
\begin{lemma}\label{lemma_relax}
The optimal $\phi_m^\cc$ to (P3-m-R) satisfies $|\phi_m^\cc|=1$.
\end{lemma}
\begin{IEEEproof}
For any solution of $\phi_m^\cc$ to (P3-m-R) with $|\phi_m^\cc|\!<\!1$, the objective value can always be increased by scaling $|\phi_m^\cc|$ to $1$, since $\beta^{\frac{1}{1\!-\!\alpha}}\!-\!\beta$ and $\beta\!-\!1$ are non-decreasing and increasing functions of $\beta$, respectively. This thus completes the proof.
\end{IEEEproof}

Lemma \ref{lemma_relax} indicates that the convex relaxation from (P3-m) to (P3-m-R) is \emph{tight}, thus the optimal solution to (P3-m-R) is also optimal for (P3-m). Thanks to the above transformations, (P3-m-R) is a convex optimization problem, whose optimal solution can be efficiently obtained via the interior-point method with complexity $\mathcal{O}(1)$. Therefore, by iteratively optimizing $(\beta,\phi_m^\cc)$ with all the other  variables $\{\phi_i^\cc,i\neq m\}_{i=1}^M$ being fixed at each time via solving (P3-m), we can obtain a feasible solution to (P3) as well as (P1), which is in general suboptimal. Note that since (P3-m) is solved optimally in every iteration, the objective value of (P3), $\beta$, is non-decreasing over the iterations, which guarantees the monotonic convergence of this algorithm since the sum-rate $r$ and hence $\beta$ is bounded above due to the finite transmit power. For each $\mv{\alpha}$, let $\tilde{r}_\II(\mv{\alpha})$ and $\tilde{r}_{\II\II}(\mv{\alpha})$ denote the obtained solutions to (P1) with $\mv{\pi}=\mv{\pi}^{\II}$ and $\mv{\pi}=\mv{\pi}^{\II\II}$, respectively. Between their corresponding rate-pairs, we further select the one with larger sum-rate as
\vspace{-2.5mm}\begin{equation} 
\!\!\!(\tilde{r}_1^\cc(\mv{\alpha}),\tilde{r}_2^\cc(\mv{\alpha}))\!\!=\!\!
\begin{cases}
(\alpha,1\!-\!\alpha)\tilde{r}_{\II}(\mv{\alpha}),\ \mathrm{if}\ \tilde{r}_\II(\mv{\alpha})\geq \tilde{r}_{\II\II}(\mv{\alpha})\\[-1mm]
(1\!-\!\alpha,\alpha)\tilde{r}_{\II\II}(\mv{\alpha}),\ \mathrm{otherwise}.
\end{cases}\!\!\!
\vspace{-2.5mm}\end{equation}
By performing time sharing among the obtained $(\tilde{r}_1^\cc(\mv{\alpha}),\tilde{r}_2^\cc(\mv{\alpha}))$'s, an inner bound of the capacity region (or an achievable rate region) is obtained as
\vspace{-2mm}\begin{equation}
\!\!\!{\mathcal{C}}_\II^\cc=\mathrm{Conv}\Big((0,0) {\bigcup}_{\mv{\alpha}:\alpha\in [0,1]}(\tilde{r}_1^\cc(\mv{\alpha}),\tilde{r}_2^\cc(\mv{\alpha}))\Big)\subseteq \mathcal{C}^\cc.
\vspace{-2mm}\end{equation}

Note that the complexity for the above proposed solution to (P1) with both decoding orders can be shown to be $\mathcal{O}(2MI)$, where $I$ denotes the number of outer iterations (each requires solving (P3-m) for $M$ times from $m\!=\!1$ to $m\!=\!M$). Therefore, by approximating the $[0,1]$ range of the rate ratio $\alpha$ with $L$ uniformly sampled points, the overall complexity for obtaining ${\mathcal{C}}_\II^\cc$ is $\mathcal{O}(2MIL)$, which is \emph{polynomial} over $M$ and thus much lower than that of \hbox{the exhaustive search (i.e., $\mathcal{O}(L_0^M)$).}
\vspace{-3mm}
\section{Capacity Region Comparison: Distributed\\[-1mm] Versus Centralized IRS Deployment}
\vspace{-2mm}
In this section, we compare the capacity regions under the two IRS deployment strategies. For simplicity, we assume that the direct user-AP channels are weak and negligible, i.e., $\bar{h}_1=\bar{h}_2=0$, which is usually the case in practical systems with severe blockage or large link distance.\footnote{The general case with non-zero $\bar{h}_1$ and $\bar{h}_2$ is more difficult to analyze, which is thus considered for the numerical example in Section \ref{sec_num}.} Moreover, for fairness, we consider the following \emph{twin} channels (defined in Assumption 1 below) between the two deployment cases, where the two distributed user-IRS channels constitute the centralized IRS-AP channel, and each user-IRS channel in the centralized case contains the corresponding IRS-AP channel in the distributed case. The twin channels hold in practice if the user-IRS distances in the distributed case are the same as the IRS-AP distance in the centralized case, and the IRS-AP distances in the distributed case are the same as the corresponding user-IRS distances in the \hbox{centralized case (see Fig. \ref{Fig_System}).}
\begin{assumption}[Twin Channels]
$\mv{g}^\cc\!=\![\mv{h}_1^{\dd^T}\!,\mv{h}_2^{\dd^T}]^T$, $h_{1m}^\cc\!=\!g_{1m}^\dd,\forall m\!\in\! \mathcal{M}_1$,  $h_{2(m+M_1)}^\cc\!=\!g_{2m}^\dd,\forall m\!\in \!\mathcal{M}_2$.
\end{assumption}
Under this assumption, we have the following proposition.
\begin{proposition}\label{prop_compare}
Under $\bar{h}_1\!\!=\!\!\bar{h}_2\!\!=\!\!0$ and Assumption 1, the capacity region of the centralized IRS deployment contains that of the distributed IRS deployment, i.e., $\mathcal{C}^\dd \subseteq \mathcal{C}^\cc$.
\end{proposition}
\begin{IEEEproof}[Proof (Sketch)]
We construct $\tilde{\mv{\Phi}}^\cc$ for the centralized IRS such that the reflection coefficients of its two sub-surfaces, $\{\tilde{\phi}_m^\cc\}_{m=1}^{M_1}$ and $\{\tilde{\phi}_m^\cc\}_{m=M_1+1}^M$,  correspond to the capacity-achieving reflection coefficients at IRS 1 and 2 for the distributed deployment shown in (\ref{eq2}), respectively, but each being rotated by a common phase $\theta_1\!\in\! [0,2\pi)$ or $\theta_2\!\in\! [0,2\pi)$, i.e.,
\vspace{-3mm}\begin{equation}\label{phase1}
\!\!\!\!\!\!\tilde{\phi}_m^\cc\!\!=\!\!
\begin{cases}
e^{j(\arg\{\bar{h}_1\}-\arg\{g_{1m}^\dd h_{1m}^\dd\}+\theta_1)},\!\!\!&m\in \mathcal{M}_1\\[-1mm] e^{j(\arg\{\bar{h}_2\}-\arg\{g_{2(m\!-\!M_1)}^\dd h_{2(m\!-\!M_1)}^\dd\}+\theta_2)},\!\!\!\!&m\!\in\! \mathcal{M}\backslash\mathcal{M}_1.\!\!\!\!\!\!
\end{cases}
\vspace{-4mm}\end{equation}
Then, we have $|\tilde{h}_k^\cc(\tilde{\mv{\Phi}}^\cc)|\!=\!|\tilde{h}_{k,\ub}^\dd\!+\!\tilde{f}_k(\theta_1,\theta_2)|$ (recall $\tilde{h}_{k,\ub}^\dd$'s are the capacity-achieving effective channel gains for the distributed case), where $\tilde{f}_k(\theta_1,\theta_2)$ is a function of $\theta_1$ and $\theta_2$. It can be shown that we can always design $\theta_1$, $\theta_2$ such that $|\tilde{h}_k^\cc(\tilde{\mv{\Phi}}^\cc)|\!\geq\!\tilde{h}_{k,\ub}^\dd$ holds for any $k\!\in\! \{1,2\}$, hence $\mathcal{C}^\dd\!\subseteq\! \mathcal{C}^\cc(\tilde{\mv{\Phi}}^\cc)\!\subseteq\! \mathcal{C}^\cc$. This thus completes the proof.
\end{IEEEproof}
\vspace{-3mm}
\section{Numerical Example}\label{sec_num}
\vspace{-2mm}
In this section, we provide a numerical example. We set $M\!\!=\!\!30$, $M_1\!\!=\!\!M_2\!\!=\!\!15$, $P_1\!\!=\!\!P_2\!\!=\!\!30$ dBm, and $\sigma^2\!\!=\!\!-90$ dBm. Under a three-dimensional coordinate system, the AP is located at $(0,0,1)$ in meter (m), and the two users are located at $(500,0,1)$ m and $(-500,0,1)$ m, respectively. The IRS in the centralized deployment is located at $(0,0,2)$ m, and the two IRSs in the distributed deployment are located at $(500,0,2)$ m and $(-500,0,2)$ m, respectively. We consider the Rayleigh fading channel model, where the entries in $\{\bar{h}_k\}$, $\{\mv{h}_k^\cc\}$ and $\{\mv{g}^\cc\}$ are generated as independent CSCG random variables with zero mean and variance equal to the path loss of the corresponding link modeled as $\gamma\!=\!\gamma_0(1/d)^{\bar{\alpha}}$, with $\gamma_0\!=\!-30$ dB, $d$ being the link distance, and $\bar{\alpha}\!=\!3$ denoting the path loss exponent. $\{\mv{h}_k^\dd\}$ and $\{\mv{g}_k^\dd\}$ are generated following the \hbox{twin channels in Assumption 1.}

\begin{figure}[t]
	\centering
	\includegraphics[width=7cm]{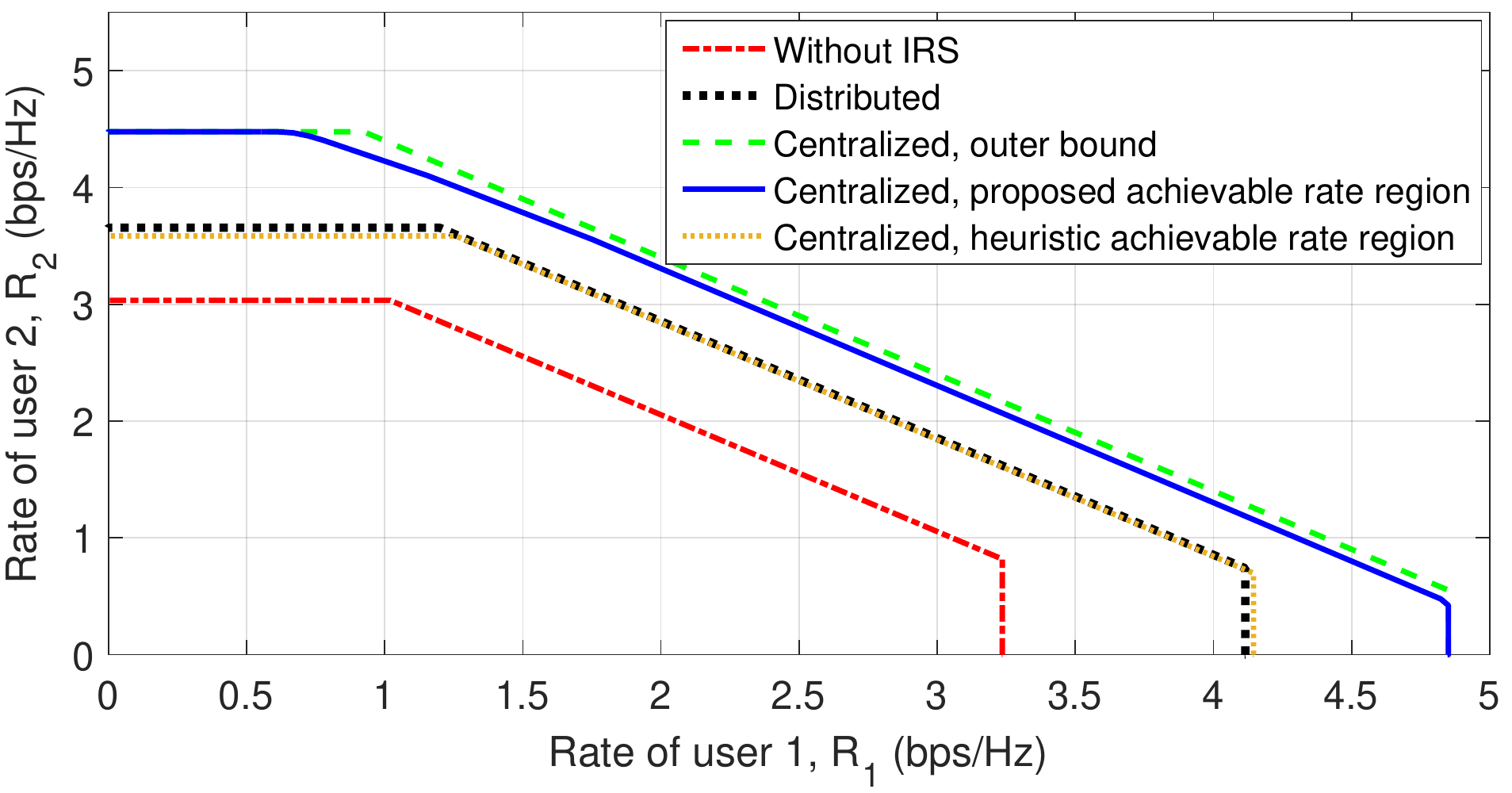}
	\vspace{-5mm}
	\caption{Capacity regions for distributed and centralized IRS deployments.}\label{Capacity_Region}
	\vspace{-7mm}
\end{figure}

In Fig. \ref{Capacity_Region}, we show the capacity region for the traditional MAC without IRS and that with two distributed IRSs, as well as the outer and inner capacity region bounds (with $L\!=\!100$) with a centralized IRS. It is observed that the capacity region inner bound for centralized deployment contains the capacity region with distributed deployment, while the latter also contains the capacity region without IRS. This thus validates the effectiveness of deploying IRS in enlarging the capacity region as well as the advantage of centralized IRS deployment over distributed deployment (even with the user-AP direct channels). It is also interesting to observe that the rate gain of centralized deployment is most pronounced when the rates of the two users are \emph{asymmetric}, since the larger passive beamforming gain provided by the centralized IRS is most useful for the user with larger rate requirement. Finally, we show the achievable rate region by a heuristic scheme under centralized deployment with $\tilde{\mv{\Phi}}^\cc$ given in (\ref{phase1}) by setting $\theta_1\!=\!\theta_2\!=\!0$ (i.e., without the additional phase rotations designed for the two sub-surfaces to further align their reflected signals). This heuristic achievable rate region is observed to be significantly smaller than our proposed one, which validates the efficacy of \hbox{our proposed rate-profile based design.}
\vspace{-2mm}\section{Conclusion}
\vspace{-2mm}
This paper studied the capacity region of an IRS-aided two-user MAC. For distributed IRS deployment, the capacity region was characterized in closed-form. For centralized IRS deployment, computationally efficient  algorithms were proposed for finding capacity region inner and outer bounds. It was revealed that centralized deployment outperforms distributed deployment under the practical channel setup, and the capacity gain is most pronounced when the user rates are asymmetric.
\vspace{-5mm}
\begin{spacing}{0.83}
\bibliographystyle{IEEEtran} 
\bibliography{IRS_Placement}

\begin{thebibliography}{10}
\providecommand{\url}[1]{#1}
\csname url@samestyle\endcsname
\providecommand{\newblock}{\relax}
\providecommand{\bibinfo}[2]{#2}
\providecommand{\BIBentrySTDinterwordspacing}{\spaceskip=0pt\relax}
\providecommand{\BIBentryALTinterwordstretchfactor}{4}
\providecommand{\BIBentryALTinterwordspacing}{\spaceskip=\fontdimen2\font plus
\BIBentryALTinterwordstretchfactor\fontdimen3\font minus
  \fontdimen4\font\relax}
\providecommand{\BIBforeignlanguage}[2]{{%
\expandafter\ifx\csname l@#1\endcsname\relax
\typeout{** WARNING: IEEEtran.bst: No hyphenation pattern has been}%
\typeout{** loaded for the language `#1'. Using the pattern for}%
\typeout{** the default language instead.}%
\else
\language=\csname l@#1\endcsname
\fi
#2}}
\providecommand{\BIBdecl}{\relax}
\BIBdecl

\bibitem{Towards}
Q.~Wu and R.~Zhang, ``{Towards smart and reconfigurable environment:
  Intelligent reflecting surface aided wireless network},'' \emph{IEEE Commun.
  Mag.}, vol.~58, no.~1, pp. 106--112, Jan. 2020.

\bibitem{Survey_Basar}
E.~Basar, M.~D. Renzo, J.~Rosny, M.~Debbah, M.-S. Alouini, and R.~Zhang,
  ``{Wireless communications through reconfigurable intelligent surfaces},''
  \emph{IEEE Access}, vol.~7, pp. 116\,753--116\,773, 2019.

\bibitem{Joint_Active}
Q.~Wu and R.~Zhang, ``{Intelligent reflecting surface enhanced wireless network
  via joint active and passive beamforming},'' \emph{IEEE Trans. Wireless
  Commun.}, vol.~18, no.~11, pp. 5394--5409, Nov. 2019.

\bibitem{Protocol}
Y.~Yang, B.~Zheng, S.~Zhang, and R.~Zhang, ``Intelligent reflecting surface
  meets {OFDM: Protocol} design and rate maximization,'' [Online]. Available:
  https://arxiv.org/abs/1906.09956.

\bibitem{ICASSP_CW}
C.~Huang, A.~Zappone, M.~Debbah, and C.~Yuen, ``{Achievable rate maximization
  by passive intelligent mirrors},'' in \emph{Proc. IEEE Int. Conf. Acoustics
  Speech Signal Process. (ICASSP)}, Apr. 2018, pp. 3714--3718.

\bibitem{Emil}
E.~Bj\"{o}rnson, {\"{O}}.~\"{O}zdogan, and E.~G. Larsson, ``{Intelligent
  reflecting surface vs. decode-and-forward: How large surfaces are needed to
  beat relaying?}'' \emph{IEEE Wireless Commun. Lett.}, vol.~9, no.~2, pp.
  244--248, Feb. 2020.

\bibitem{CE_Johansson}
D.~Mishra and H.~Johansson, ``{Channel estimation and low-complexity
  beamforming design for passive intelligent surface-assisted MISO wireless
  energy transfer},'' in \emph{Proc. IEEE Int. Conf. Acoustics Speech Signal
  Process. (ICASSP)}, May 2019.

\bibitem{CE_B}
B.~Zheng and R.~Zhang, ``{Intelligent reflecting surface-enhanced OFDM: Channel
  estimation and reflection optimization},'' \emph{IEEE Wireless Commun.
  Lett.}, {Early Access}.

\bibitem{MIMO}
S.~Zhang and R.~Zhang, ``Capacity characterization for intelligent reflecting
  surface aided {MIMO} communication,'' {\emph{IEEE J. Sel. Areas Commun.}}, to
  appear. [Online]. Available: https://arxiv.org/abs/1910.01573.

\bibitem{Elements}
T.~M. Cover and J.~A. Thomas, \emph{{Elements of Information Theory}}.\hskip
  1em plus 0.5em minus 0.4em\relax Wiley, 2006.

\bibitem{SDR}
Z.-Q. Luo, W.-K. Ma, A.~M.-C. So, Y.~Ye, and S.~Zhang, ``{Semidefinite
  relaxation of quadratic optimization problems},'' \emph{IEEE Signal Process.
  Mag.}, vol.~27, no.~3, pp. 20--34, May 2010.

\bibitem{Cooperative}
R.~Zhang and S.~Cui, ``{Cooperative interference management with MISO
  beamforming},'' \emph{IEEE Trans. Signal Process.}, vol.~58, no.~10, pp.
  5450--5458, Oct. 2010.

\end{thebibliography}
\end{spacing}
\end{document}